\documentstyle[prb, preprint,dcolumn, epsfig, aps ]{revtex}
\usepackage{graphicx}

\begin{document}
\begin{frontmatter}
\title{Insulator superconductor transition on solid inert gas substrates}

\author{K. Das Gupta}
\author{Swati S. Soman}
\author{G. Sambandamurthy} 
\author{N. Chandrasekhar\thanksref{thank1}}
\address{ Department of Physics, Indian Institute of Science, 
Bangalore, India}

\thanks[thank1]{Corresponding author. 
 E-mail: chandra@physics.iisc.ernet.in}

\begin{abstract}
We present observations of the insulator-superconductor transition in ultrathin films of Bi on solid xenon condensed on quartz and on Ge on quartz. The relative permeability $\epsilon_{r}$ ranges from
1.5 for Xe to 15 for Ge. Though we find screening effects as expected, the I-S transition is robust, and
unmodified by the substrate. The resistance separatrix is found to be close to $h/4e^2$ and the crossover thickness
close to 25 $\AA$ for all substrates. I-V studies and Aslamazov-Larkin analyses indicate superconductivity is
inhomogeneous. The transition is best described in terms of a percolation model. 
\end{abstract}

\end{frontmatter}

\section{Introduction}
The insulator-superconductor (I-S) transition has been
extensively investigated over the last decade, in a variety of
systems such as thin films,~\cite{hav,gra} single Josephson junctions,~\cite{pen}
arrays,~\cite{zan} and one-dimensional wires.~\cite{tin} Values of limiting resistance
close to the quantum resistance for pairs in one dimensional
wires and two-dimensional nominally homogeneous
ultrathin films are reported. Differing values of the
limiting resistance at the transition have been observed~\cite{gra} in
different systems, and attributed to structure, i.e., homogeneous
and granular films are expected to behave differently.
A phase-only picture, first proposed by Ramakrishnan~\cite{tvr}
and further elaborated by Fisher~\cite{fis} has been considered appropriate
for such systems. A scaling theory of the I-S transition
has been developed.~\cite{fis2}

\section{Experiments and Results}
Fig. 1 shows the evolution of the temperature dependence
of the sheet resistance R(T) with thickness for  Bi
films on Ge 10 ${\mathrm \AA}$ thick, which has been deposited on amorphous
quartz. A insulator-superconductor transition is immediately apparent.
A similar result is obtained for Bi films on solid
xenon condensed on amorphous quartz, as shown in fig. 2. A transition from
insulating type behavior, to superconducting behavior as the
thickness of the films is increased is clear. This type of zero
field transition is considered a zero temperature quantum
phase transition, controlled either by disorder, carrier concentration,
or thickness. The normal state resistance at an
arbitrarily high temperature $R_{\mathrm N}$ has traditionally been used to
parametrize the transition, although it may be weakly temperature
dependent above the superconducting transition
temperature, and becomes ill defined as the I-S transition is
approached. The value of the normal state resistance of a
film on the boundary between superconducting and insulating
behavior has been referred to as the resistance separatrix,
and has been denoted by $R_{\mathrm 0}$.~\cite{hav,gra,tin} 
We obtain $R_{\mathrm 0}$ as an algebraic
average of the sheet resistances of the last insulating
and the first superconducting films, measured at a relatively
high temperature ~10 K. $R_{\mathrm 0}$ is close to $h/4e^2$ for both
sets of data. This observation indicates that the value of $R_{\mathrm 0}$ is
substrate independent, and possibly experiment independent.

\begin{figure}[t]
\begin{center}\leavevmode
\includegraphics[width=0.8\linewidth]{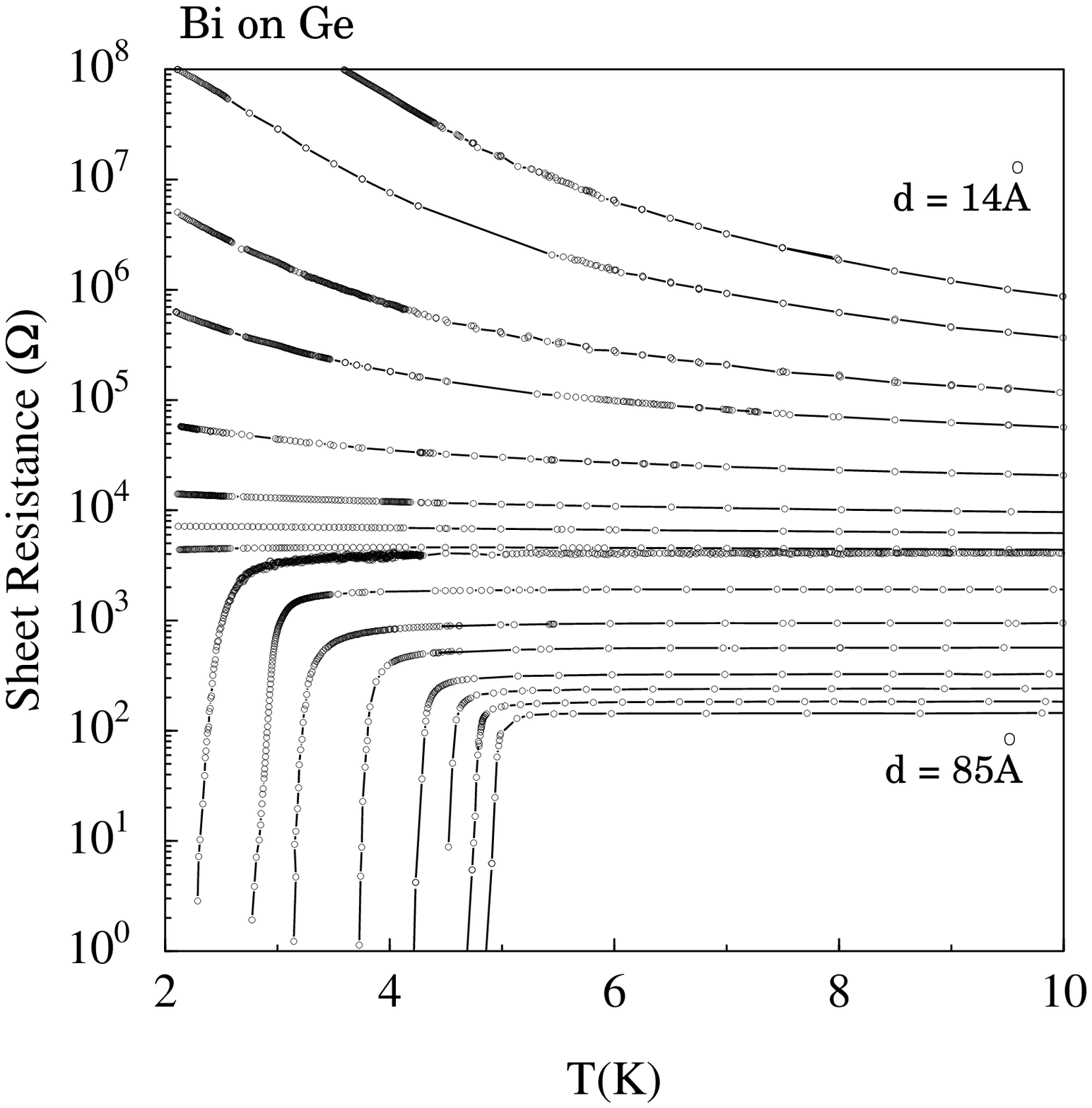}
\caption{ IST of Bi on Ge underlayer.
}\label{Fig. 1}\end{center}\end{figure}

\begin{figure}[b]
\begin{center}\leavevmode
\includegraphics[width=0.8\linewidth]{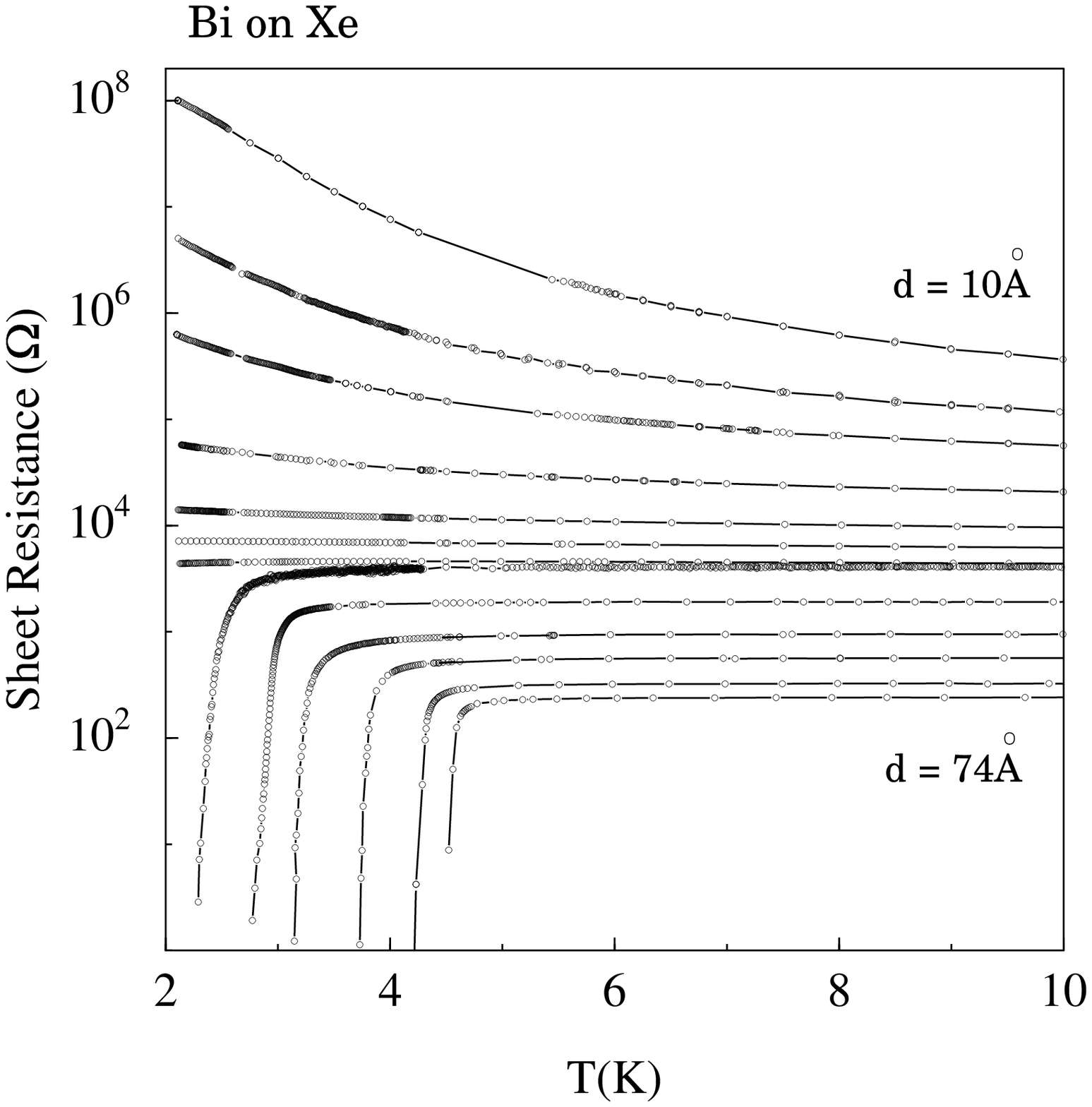}
\caption{ IST of Bi on Xe underlayer.
}\label{Fig. 2}\end{center}\end{figure}

The form of the $R(T)$ for these films may lead us to the
conclusion that these films are nominally homogeneous.
Such a conclusion is incorrect as we show below. 
We find that superconductivity
in our films is indeed percolative in nature~\cite{gsm}. Aslamazov
and Larkin~\cite{al} considered the possibility of fluctuations causing
superconductivity. The total conductivity is given by 
$\sigma = \sigma_{\mathrm N}+ \sigma'_{\mathrm 2D}$, 
where $\sigma_{\mathrm N}$ is the normal state dc conductivity,
and $\sigma'_{\mathrm 2D}$ the paraconductivity. Its temperature dependance is
similar to that of the magnetic susceptibility at $T-T_{\mathrm c}$ . They
derived the result
\begin{equation} 
\frac{\sigma'_{\mathrm 2D} }{\sigma_{\mathrm N}}=
\frac{e^{2}}{16 \hbar}\frac{R^{\mathrm N}_{\Box}}{\tau}. 
\end{equation}
where $R^{\mathrm N}_{\Box}$ is the normal state sheet resistance and 
$\tau = (T-T_{\mathrm c})/T_{\mathrm c}$ is the width factor. 
$T_{\mathrm c}$ is
the mean field transition temperature. 
$\tau/R^{\mathrm N}_{\Box}=g_{\mathrm AL}=
\frac{e^{2}}{16 \hbar}$ is a constant
for all materials. We have evaluated
$\tau/R^{\mathrm N}_{\Box}=g_{\mathrm exp}$ for various films. 
A systematic dependence of
gexp on the thickness d is shown in Fig. 3. This parameter
deviates from $g_{\mathrm AL}$ for thinner films. It approaches the AL
value as the thickness is increased. It is assumed that
theory predicts the same $g_{\mathrm AL}$ for all films, 
independent of microstructure.
Both the normal state conductance and paraconductance
depend on sample shape. Glover~\cite{glo} has shown that
as the microstructure deviates from a ''uniform rectangular
slab,'' gexp exceeds the AL value. The thinner the film,
higher the disorder, larger are the deviations from a slab
geometry, and larger the deviation of gexp from the AL value.
Hence, films close to the transition are inhomogeneous.

\begin{figure}[t]
\begin{center}\leavevmode
\includegraphics[width=0.8\linewidth]{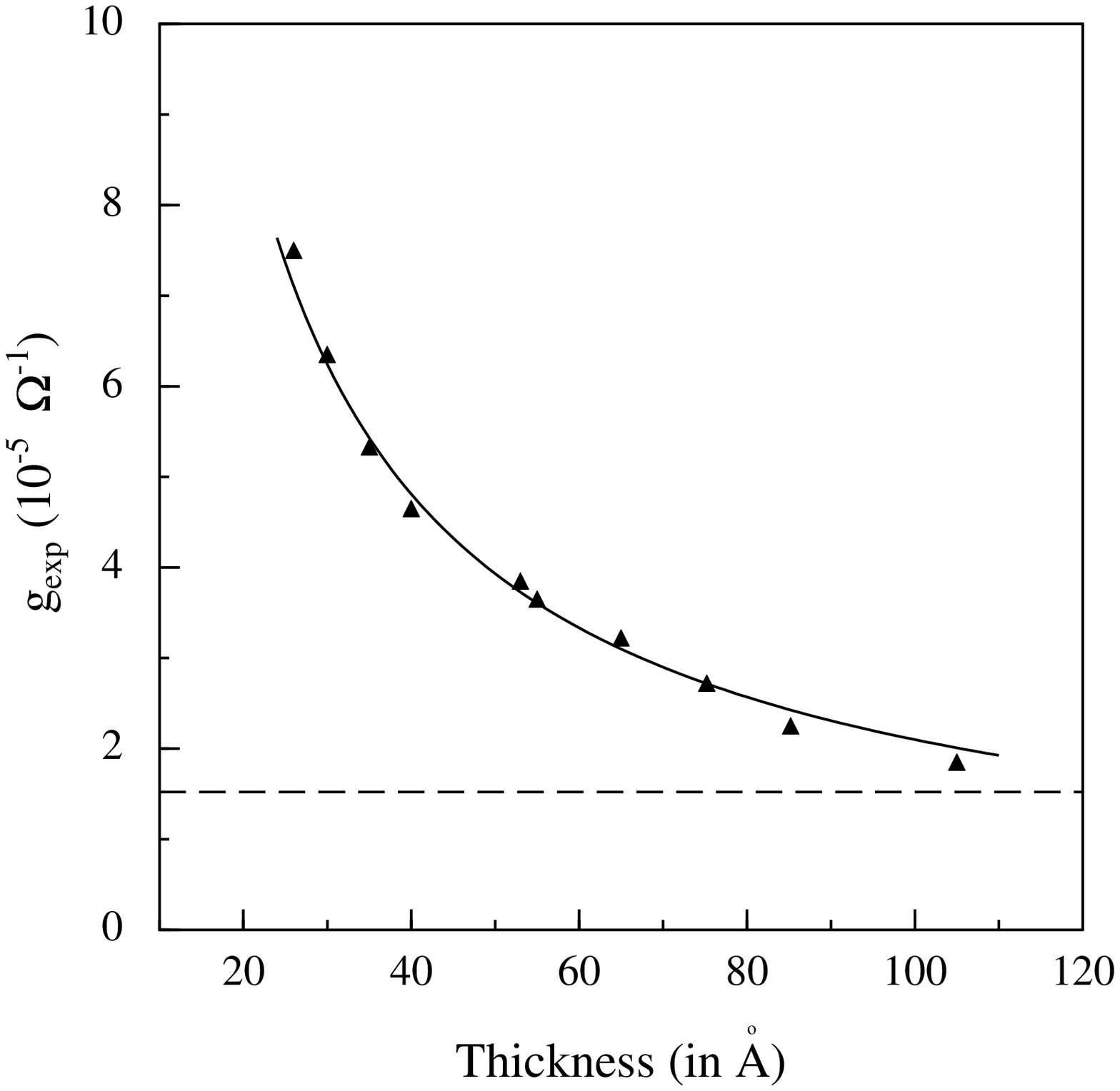}
\caption{ Variation of the AL parameter with thickness.
}\label{Fig. 3}\end{center}\end{figure}

\begin{ack}
The work is supported by DST and UGC, Government of India. KDG thanks CSIR for the research fellowship.
\end{ack}

\end{document}